\documentclass{moriond}

\def\be{\begin{equation}}
\def\ee{\end{equation}}
\def\bea{\begin{eqnarray}}
\def\eea{\end{eqnarray}}

\newcommand{\Photo}{}

\begin{document}
\vspace*{4cm}
\title{Two-Photon Fusion Results at BESIII}

\author{ M. Lellmann on behalf of the BESIII collaboration }

\address{Johannes Gutenberg University, Johann-Joachim-Becher-Weg 45, D-55099 Mainz, Germany}

\maketitle\abstracts{
Two-photon processes in electron-positron collisions of the form $e^+e^-\to e^+e^-X$ are a well-established tool for probing hadronic structure and provide a direct access to non-vector states in $e^+e^-$ collisions. Collisions of quasi real photons have been employed by several experimental collaborations to extract radiative widths of various hadronic resonances.
Persistent discrepancies between Standard Model predictions and measurements of the muon anomalous magnetic moment $a_\mu = (g-2)\mu/2$ have renewed interest in theoretical predictions for $\gamma\gamma \to X$ processes, particularly with off-shell photons. These are essential for evaluating the hadronic light-by-light contribution to $a_\mu$.
This work reviews results and outlines future two-photon measurements by the BESIII collaboration.
}

\section{Introduction}
The two-photon production of hadrons in electron-positron scattering, $e^+e^- \to e^+e^-X$, where the hadronic state $X$ is formed via the fusion of two photons, offers a clean environment for studying non-vector states at $e^+e^-$ colliders. Photon fusion allows the direct production of states with quantum numbers $J^{PC} = 0^{\pm+},\, 2^{\pm+},\, \dots$, and, when at least one photon is highly virtual, also permits the production of axial-vector states.

Two-photon interactions are purely electromagnetic, resulting in suppressed gluonium production, and thus can serve as anti-gluon filters.

The cross-section of hadron production via two-photon fusion is directly related to the $e^+e^- \to e^+e^-X$ process~\cite{Budnev:1975poe,Pascalutsa:2012pr}, and depends on three Lorentz invariants: the two-photon invariant mass $W$ and the photon virtualities $Q_{1,2}^2 = -(p_{1,2} - p^\prime_{1,2})^2$, with $p_{1,2}$ and $p^\prime_{1,2}$ denoting the initial and final four-momenta of the electron and positron. Additional angular variables in the hadronic system may enter when multiple hadrons are produced. 

Cross-sections typically decrease with increasing photon virtualities, peaking at $Q_1^2 = Q_2^2 \approx 0$. This kinematic regime, which can be studied when both final-state leptons escape at small angles ($Q^2=-(p-p^\prime)\approx 4EE^\prime\sin^2\theta/2$), is particularly useful for light meson spectroscopy. The relatively high statistics in combination with the direct dependence of the cross-sections on the radiative widths provide essential information of the densely populated spectra in the light meson regime, which is challenging to describe theoretically. Such studies advance the understanding of light meson structure and help scrutinize glueball candidates.

When studying events with highly virtual photons, which in turn means measuring at least one of the final state leptons at large scattering angles, one has the opportunity to study the electromagnetic transition form factors of the produced resonances in dependence on the photon virtuality. This quantity describes the coupling of photons to hadronic matter and provides structural information on the hadrons. The information is of large interest for the Standard Model prediction of the hadronic Light-by-Light scattering (HLbL) contribution to the muon anomalous magnetic moment $a_\mu=(g-2)_\mu/2$~\cite{Danilkin:2019mhd}, where a long standing discrepancy between standard model and direct measurement is observed \cite{Aoyama:2020ynm,PhysRevD.110.032009}. Especially the study of the two-photon production of pseudoscalars, pion pairs, and axial mesons at photon virtualities in the order of 1\,GeV$^2$ are of utmost importance\cite{Danilkin:2019mhd}.

In the following, key measurements of the BESIII collaborations of two-photon processes for both, light meson spectroscopy and experimental input for $a_\mu$, are presented and future measurements by the BESIII collaboration are discussed.

\section{The BESIII Experiment}
The BESIII detector~\cite{BESIII:2009fln} records symmetric $e^+e^-$ collisions 
provided by the BEPCII storage ring~\cite{Yu:IPAC2016-TUYA01}
in the center-of-mass energy range from 1.84 to 4.95~GeV,
with a peak luminosity of $1.1 \times 10^{33}\;\mathrm{cm}^{-2}\mathrm{s}^{-1}$ 
achieved at $\sqrt{s} = 3.773\;\mathrm{GeV}$. 
BESIII has collected large data samples in this energy region~\cite{Ablikim:2019hff,EventFilter,EcmsMea}. The cylindrical core of the BESIII detector covers 93\% of the full solid angle and consists of a helium-based
multilayer drift chamber~(MDC), a time-of-flight
system~(TOF), and a CsI(Tl) electromagnetic calorimeter~(EMC),
which are all enclosed in a superconducting solenoidal magnet
providing a 1.0~T magnetic field.
The solenoid is supported by an
octagonal flux-return yoke with resistive plate counter muon
identification modules interleaved with steel. 
The charged-particle momentum resolution at $1~{\rm GeV}/c$ is
$0.5\%$, and the 
${\rm d}E/{\rm d}x$
resolution is $6\%$ for electrons
from Bhabha scattering. The EMC measures photon energies with a
resolution of $2.5\%$ ($5\%$) at $1$~GeV in the barrel (end cap)
region. The time resolution in the plastic scintillator TOF barrel region is 68~ps, while
that in the end cap region was 110~ps. The end cap TOF system was upgraded in 2015 using multigap resistive plate chamber technology, providing a time resolution of 60~ps~\cite{etof1,etof2,etof3}.

\section{Study $\pi^0\pi^0$, $\pi^0\eta$, and $K^+K^-$ Production from Real Photon Collisions}
\begin{figure}[b!]
	\centering
	\includegraphics[width=1.0\linewidth]{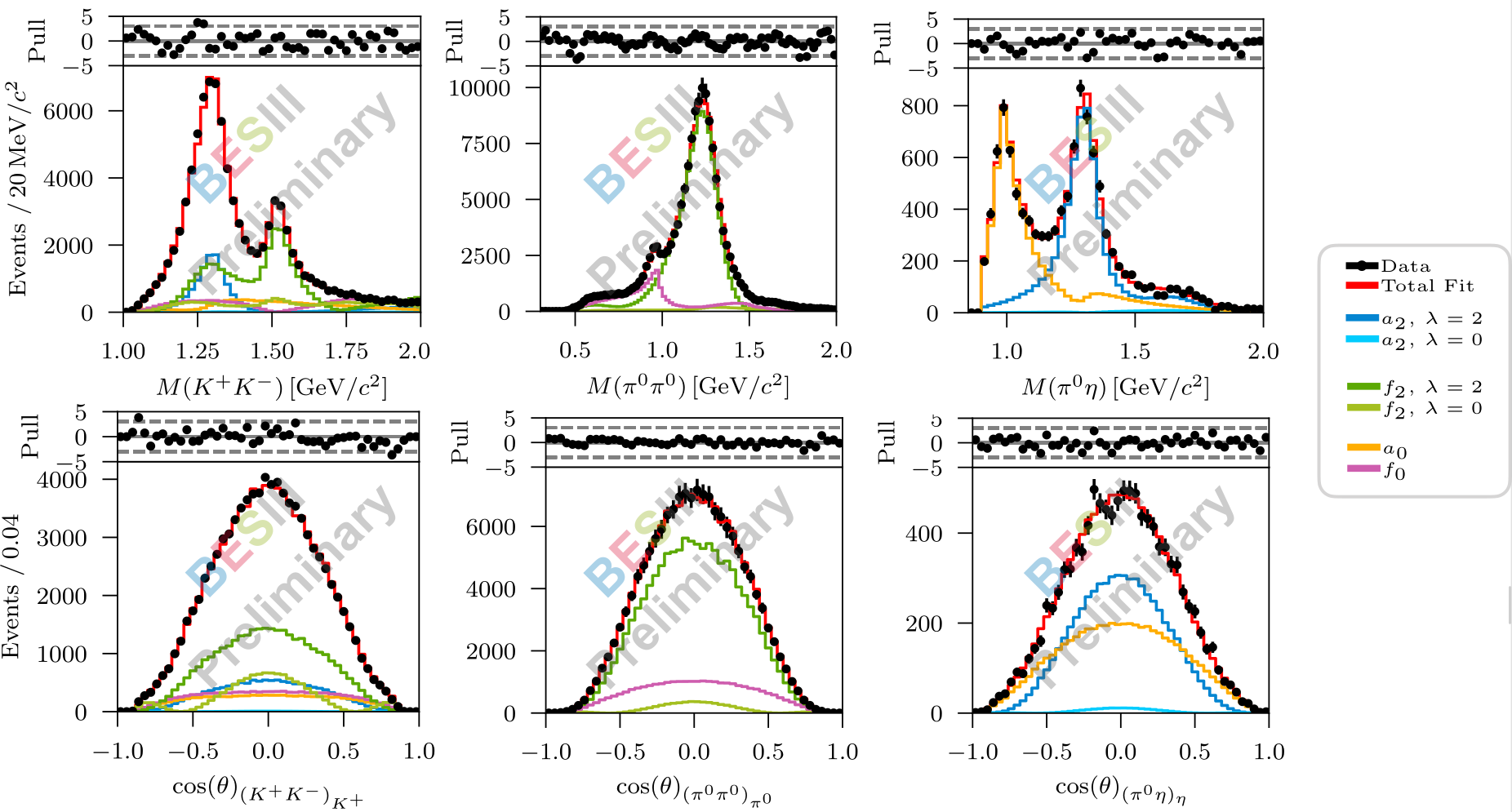}
	\caption{Individual contributions from the best fit to the $K^+K^-$ invariant mass and angular distributions (left panels), $\pi^0\pi^0$ (middle), and $\pi^0\eta$ (right). The uncertainties are statistical and derived using bootstrapping.}
	\label{fig:pwa}
\end{figure}
The BESIII collaboration measured the production of $\pi^0\pi^0$, $\pi^0\eta$, and $K^+K^-$ in the collision of two quasi-real photons and, for the first time, performed a coupled-channel partial wave analysis on the results \cite{Kuessner2022}. The analysis is based on data collected at center-of-mass energies between $\sqrt{s} = 3.7$ and 4.7\,GeV, corresponding to a total integrated luminosity of 21.3\,fb$^{-1}$. High-purity samples were obtained using event-based background subtraction techniques that exploit differences in the transverse momentum distributions of signal and background. The data were interpreted using the K-matrix formalism with a P-vector approach, with resonance parameters - except for the radiative widths - fixed from previous analyses of $p\bar{p}$ and scattering data \cite{CrystalBarrel:2019zqh,Kopf:2020yoa}, resulting in the most precise measurements to date of the radiative widths for the $f_2(1270)$ and $a_2(1320)$, the first determination of helicity contributions to the $f_2^\prime(1525)$, and the first measurements of the radiative widths of the $f_0(1370)$, $f_0(1500)$, and $f_0(1710)$. The best fit result is displayed together with the data in Fig.~\ref{fig:pwa}. Using the results of the partial wave analysis, the extracted cross-sections are interpolated to the full angular range, providing valuable input for the determination of HLbL contribution to $a_\mu$ ~\cite{Danilkin:2019mhd,Kuessner2022}. The publication of the final results of this analysis is foreseen in the near future.

\section{Study of the $\gamma\gamma^\ast\to\pi^0$ Transition Form Factor}
\begin{figure}[h!bt]
	\centering
	\includegraphics[width=0.6\linewidth]{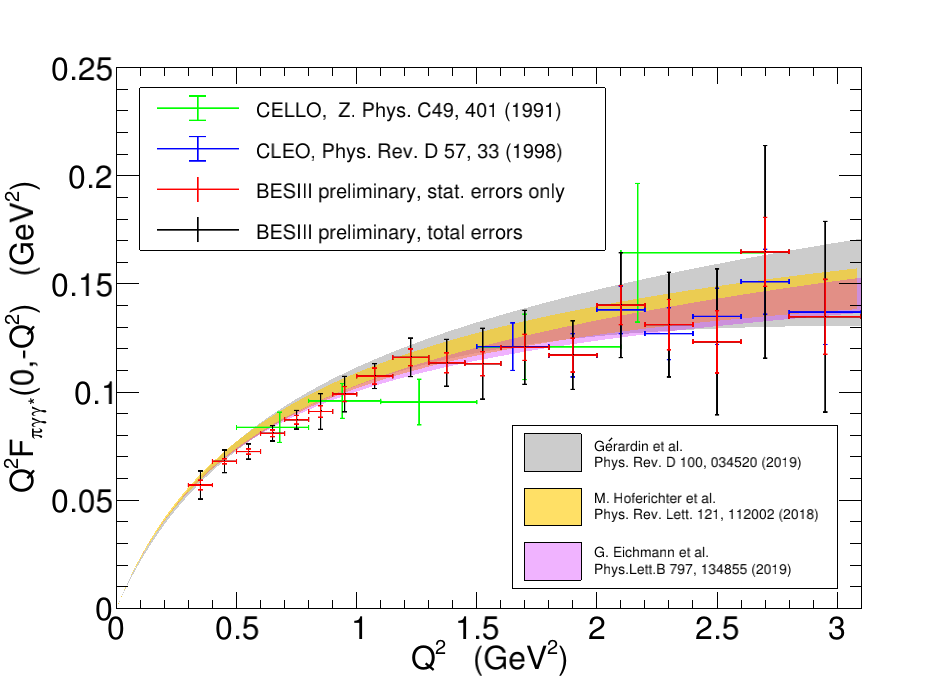}
    \caption{ Comparison of the measured $\pi^0$ transition form factor to CELLO and CLEO results as well as phenomenological and lattice QCD calculations. }
	\label{fig:tff}
\end{figure}

BESIII measured the $\gamma\gamma^* \to \pi^0$ transition form factor (TFF), a key input for calculating $a_\mu^\mathrm{HLbL}$. The analysis uses 2.9\,fb$^{-1}$ of data collected at $\sqrt{s} = 3.773$\,GeV, selecting events where one final-state lepton and the pion decay products are detected. Backgrounds are suppressed through kinematic cuts, which also constrain the undetected lepton to small angles, and finally subtracted using data driven methods.
The TFF is extracted from the $Q^2$-dependent differential cross-section by normalizing to the theoretical cross-section with constant TFF taken from the \textsc{Ekhara}~3.0 generator~\cite{Czyz:2018jpp}. As shown in Fig.~\ref{fig:tff}, this is the first measurement for $Q^2 < 0.5$\,GeV$^2$, the most precise below 2\,GeV$^2$, and remains competitive at higher $Q^2$ \cite{CELLO:1990klc,CLEO:1997fho}. The results are consistent with both phenomenological\cite{Hoferichter:2018dmo,Eichmann:2019tjk} and lattice QCD\cite{Gerardin:2019vio} predictions. A final version of the measurement, including radiative corrections and corrections related to the residual virtuality of the pseudo-real photon, will be published shortly. A new measurement with 20.3\,fb$^{-1}$ of data is ongoing.

\section{Conclusion and Outlook}
Two-photon interactions are a valuable probe of light meson structure and offer critical input for addressing the muon anomaly. BESIII provides ideal experimental conditions for measuring meson radiative widths and for studying two-photon transition form factors in the kinematic regime relevant to precision calculations of $a_\mu^\mathrm{HLbL}$.

Several future measurements are planned. Studies of singly virtual production of pion pairs and $\pi\eta$ final states in a similar kinematic range as the presented $\pi^0$ results are well advanced. These will include, for the first time, the measurement of $\gamma\gamma^* \to \pi^+\pi^-$. For neutral pion pairs, advanced event reconstruction techniques will enable, for the first time, extraction of the doubly virtual production cross-section in a limited kinematic region, as well as studies of interference cross-sections via angular modulations in the two-photon center-of-mass system. The latter requires a newly developed Monte Carlo event generator capable of numerically evaluating the relevant QED couplings. This generator also includes a phenomenological model for the $\gamma\gamma^\ast \to f_1(1285) \to \eta\pi^+\pi^-$ process\cite{Ren:2024uui}, which is currently under experimental investigation. Besides the singly virtual measurements, first studies of the double virtual production of the light pseudoscalar mesons are performed. All upcoming analyses will benefit from a significantly larger data set collected at a single energy point $\sqrt{s} = 3.773$\,GeV with an integrated luminosity of 20.3\,fb$^{-1}$.

\section*{Acknowledgments}
This work was supported by the Deutsche Forschungsgemeinschaft (DFG, German Research Foundation) through the Research Unit FOR 5327 (Photon-photon interactions in the Standard Model and beyond, Projektnummer 458854507).

\section*{References}
\bibliography{moriond}

\end{document}